\documentclass[letterpaper]{article}
\usepackage[usenames,dvipsnames]{color}
                           
\usepackage{aaai}
\usepackage{times}
\usepackage{helvet}
\usepackage{courier}
\usepackage{graphicx}
\usepackage{multirow}
\usepackage{arydshln} 

\usepackage{tikz}
\usetikzlibrary{decorations.pathreplacing,calc}

\usepackage{amsmath}
\usepackage{amssymb}
\usepackage{ stmaryrd }
\usepackage[ruled,vlined,linesnumbered]{algorithm2e}
\usepackage{ltlfonts}

\begin{document}

%
 \title{From Realizability Modulo Theories to Synthesis Modulo Theories \\
 Part 1: Dynamic approach}
\author{Andoni Rodríguez and César Sánchez \\
    IMDEA Software Institute, Madrid. Spain \\
    Universidad Politécnica de Madrid. Spain}

\maketitle

\newcommand{\Nat}{\mathbb{N}}
\newcommand{\KWD}[1]{\textit{#1}}
\newcommand{\topoly}{\KWD{topoly}}
\newcommand{\inside}{\KWD{inside}}
\newcommand{\outside}{\KWD{outside}}
\newcommand{\toconf}{\KWD{toconf}}
\newcommand{\True}{\KWD{true}}
\newcommand{\False}{\KWD{false}}
\newcommand{\BL}{\KWD{bl}}
\newcommand{\DefinedAs}{\,\stackrel{\text{def}}{=}\,}
\newcommand{\Confs}{\KWD{Confs}}


\newcommand{\Theo}{\mathcal{T}}
\newcommand{\Boolbb}{\mathbb{B}}
\newcommand{\PhiB}{\varphi^{\Boolbb}}
\newcommand{\PhiT}{\varphi^{\Theo}}
\newcommand{\PhiExtra}{\varphi_{\textit{Extra}}}
\newcommand{\Sys}{\mathcal{S}}
\newcommand{\Env}{\mathcal{E}}
\newcommand{\sysStratT}{\rho_{\Sys}^{\Theo}}
\newcommand{\sysStratB}{\rho_{\Sys}^{\Boolbb}}
\newcommand{\envStratT}{\rho_{\Sys}^{\Theo}}
\newcommand{\envStratB}{\rho_{\Env}^{\Boolbb}}
\newcommand{\machinT}{\mathcal{M}^{\Theo}}
\newcommand{\machinB}{\mathcal{M}^{\Boolbb}}
\newcommand{\deltaT}{\delta_{\Theo}}
\newcommand{\deltaB}{\delta_{\Boolbb}}
\newcommand{\outpT}{o_{\Theo}}
\newcommand{\outpB}{o_{\Boolbb}}
\newcommand{\sImplies}{\shortrightarrow}
\newcommand{\barX}{\overline{x}}
\newcommand{\barY}{\overline{y}}
\newcommand{\barV}{\overline{v}}
\newcommand{\richY}{\mathcal{Y}^{\textit{LTL}_{\Theo}}}
\newcommand{\suchThat}{\textit{. }}

\newcommand{\ThA}{\Theo_{\mathbb{A}}}
\newcommand{\LTL}{\ensuremath{\textit{LTL}}\xspace}
\newcommand{\LTLt}{\ensuremath{\textit{LTL}_{\Theo}}\xspace}
\newcommand{\naturalNums}{\mathbb{N}}
\newcommand{\integerNums}{\mathbb{Z}}
\newcommand{\rationalNums}{\mathbb{Q}}
\newcommand{\realNums}{\mathbb{R}}
\newcommand{\complexNums}{\mathbb{C}}
\newcommand{\ThN}{\Theo_{\naturalNums}}
\newcommand{\ThQ}{\Theo_{\rationalNums}}
\newcommand{\ThC}{\Theo_{\complexNums}}
\newcommand{\ThAPF}{\Theo_{\textit{APF}}}
\newcommand{\ThSeq}{\Theo_{\textit{Seq}}}
\newcommand{\ThT}{\Theo_{\mathbb{T}}}

\newcommand{\cesartodo}[1]{\todo[linecolor=blue,backgroundcolor=blue!25,bordercolor=blue]{Cesar: #1}}
\newcommand{\cesartext}[1]{\todo[inline,linecolor=blue,backgroundcolor=blue!25,bordercolor=blue]{Cesar: #1}}
\newcommand{\andonitodo}[1]{\todo[linecolor=red,backgroundcolor=red!25,bordercolor=red]{Andoni: #1}}
\newcommand{\andonitext}[1]{\todo[inline,linecolor=red,backgroundcolor=red!25,bordercolor=red]{Andoni: #1}}

\newcommand{\AP}{\ensuremath{\mathsf{AP}}\xspace}
\newcommand{\alphabet}{\mathrm{\Sigma}}
\newcommand{\DefOR}{\ensuremath{\hspace{0.2em}\big|\hspace{0.2em}}}

\newcommand{\Always}{\LTLsquare}
\newcommand{\Event}{\LTLdiamond} 
\newcommand{\Next}{\LTLcircle}
\newcommand{\LTLPrev}{\LTLcircleminus}
\newcommand{\PrevNoFirst}{\LTLcircletilde}
\newcommand{\HasAlwaysBeen}{\LTLsquareminus}
\newcommand{\Once}{\LTLdiamondminus}
\newcommand{\Since}{\mathbin{\mathcal{S}}}
\newcommand{\BackTo}{\mathbin{\mathcal{B}}}
\newcommand{\WeakPrev}{\LTLcircletilde}
\newcommand{\U}{\mathbin{\mathcal{U}}}
\newcommand{\R}{\mathbin{\mathcal{R}}}

\newcommand{\Not}{\mathclose{\neg}}
\newcommand{\Or}{\mathrel{\vee}}
\newcommand{\Impl}{\mathrel{\rightarrow}}
\newcommand{\Into}{\Impl}


\newcommand{\configuration}{choice\xspace}
\newcommand{\configurations}{choices\xspace}
\newcommand{\Configuration}{Choice\xspace}
\newcommand{\Configurations}{Choices\xspace}

\newcommand{\reaction}{reaction\xspace}
\newcommand{\reactions}{reactions\xspace}
\newcommand{\xbar}{\overline{x}}
\newcommand{\ybar}{\overline{y}}

\newcommand{\Lit}{\ensuremath{\mathit{Lit}}\xspace}
\newcommand{\phiT}{\ensuremath{\varphi_{\mathcal{T}}}\xspace}
\newcommand{\phiB}{\ensuremath{\varphi_{\mathbb{B}}}\xspace}

\newcommand{\Myphi}[1]{\ensuremath{\varphi^{\textit{#1}}}}
\newcommand{\myphi}[1]{\Myphi{#1}\xspace}
\newcommand{\phiLegal}{\myphi{legal}}
\newcommand{\phiExtra}{\myphi{extra}}
\newcommand{\phiExtraP}{\ensuremath{{\varphi^{\textit{extra}}}'\xspace}}
\newcommand{\phiEx}{\ensuremath{\phiExtra}\xspace}

\newcommand{\phiAddT}{\ensuremath{\Myphi{add}_{\mathcal{T}}}\xspace}
\newcommand{\phiAdd}{\myphi{add}}

\newcommand{\phiCar}{\myphi{car}}

\newcommand{\xs}{\ensuremath{\overline{x}}\xspace}
\newcommand{\ys}{\ensuremath{\overline{y}}\xspace}
\newcommand{\vs}{\ensuremath{\overline{v}}\xspace}
\newcommand{\ubar}{\ensuremath{\overline{u}}\xspace}
\newcommand{\Pt}{\ensuremath{\mathit{pt}}}
\newcommand{\Nt}{\ensuremath{\mathit{nt}}}
\newcommand{\DEC}{\textit{dec}}
\newcommand{\PT}[1]{#1^p}
\newcommand{\NT}[1]{#1^a}

\newcommand{\mycal}[1]{\ensuremath{\mathcal{#1}}\xspace}
\newcommand{\calC}{\mycal{C}}
\newcommand{\react}{\ensuremath{\textit{react}}}
\newcommand{\calR}{\mycal{R}}
\newcommand{\calT}{\mycal{T}}
\newcommand{\calQ}{\mycal{Q}}
\newcommand{\VR}{\ensuremath{\textit{VR}}\xspace}
\newcommand{\bconf}{\ensuremath{\textit{bconf}}\xspace}

\newcommand{\VE}{\ensuremath{\overline{v}_e}}
\newcommand{\VS}{\ensuremath{\overline{v}_s}}
\newcommand{\LoopStop}{\textit{loop\_stop}}
\newcommand{\GetReaction}{\textit{get\_react}}
\newcommand{\ValidityReduce}{\textit{validity\_and\_reduce}}
\newcommand{\ThZ}{\mathcal{T}_\mathbb{Z}}
\newcommand{\ThR}{\mathcal{T}_\mathbb{R}}

\newcommand{\GameT}{\mathcal{G}^{\calT}}
\newcommand{\GameB}{\mathcal{G}^{\mathbb{B}}}

\newcommand{\qrprec}{\mathrel{\preceq}}

\newcommand{\qreact}{\ensuremath{\textit{qreact}}}
\newcommand{\Vars}{\mathit{Vars}}
\newcommand{\Bool}{\mathbb{B}}
\newcommand{\Part}{\rightharpoonup}

\newcommand{\LTLf}{$\textrm{LTL}_f$ }




\newcommand{\TableBenchmark}{

\begin{table*}[t!]
 \centering
\begin{tabular}{|c|c||c|c|c|c||c|c|c|c|c|c|c|c|c|}  \hline
  Bn. & Cls. & \multicolumn{4}{|c|}{\textit{BA}} & 
  \multicolumn{3}{|c||}{$1K$ sims.} & \multicolumn{3}{|c||}{$10K$ sims.} & \multicolumn{3}{|c|}{$10K$ sims. (sft.)}\\ 
 \cline{3-15}
  (nm.) & (vr, lt) & Tme. & Quer. & Dc. & $\mathbb{B}$. & 
  $\textit{Pa.}$ & $\textit{Cn.}$ & $\textit{Pr.}$ & $\textit{Pa.}$ & $\textit{Cn.}$ & $\textit{Pr.}$ & $\textit{m/m.}$ & $\textit{pc.}$ & $\textit{sm.}$ \\ [0.5ex] 
  
 \hline \hline
 
\multirow{4}{*}{\textit{Li.}} &  (1, 7) & 28.66 & 1008 & 1 & \multirow{4}{*}{3.71} & 
0.01 & \multirow{4}{*}{2.17} & \multirow{4}{*}{240} & 0.01 & \multirow{4}{*}{2.14} & \multirow{4}{*}{2.32} & \multirow{4}{*}{216} & \multirow{4}{*}{204} & \multirow{4}{*}{202} \\ 
 &  (2, 4) & 0.72 & 44 & 16 & & 
 0.01 & & & 0.01 & & & & & \\
 &  (1, 3) & 0.51 & 30 & 4 & & 
 0.01 & & & 0.01 & & & & & \\
 &  (1, 2) & 0.13 & 7 & 3 & & 
 0.01 & & & 0.01 & & & & & \\
 
 \hline \hline
 
  \multirow{8}{*}{\textit{Tr.}} &  (1, 3) & 0.87 & 45 &  5 & \multirow{8}{*}{5.01} & 
  0.01 & \multirow{8}{*}{3.41} & \multirow{8}{*}{272} & 0.01 & \multirow{8}{*}{3.39} &  \multirow{8}{*}{262} & \multirow{8}{*}{294} & \multirow{8}{*}{270} & \multirow{8}{*}{306}\\ 
 &  (2, 1) & 0.04 & 2 & 2 &  & 
 0.01 & & & 0.01 & & & & & \\
 &  (1, 3) & 0.19 & 12 & 9 & & 
 0.01 & & & 0.01 & & & & & \\
 & (1, 1) & 0.10 & 5 & 4 &  & 
 0.01 & & & 0.01 & & & & & \\
    & (3, 6) & 104.5 & 5367 & 15 &  & 
    0.02 & & & 0.01 & & & & & \\ 
 &  (4, 5) & 3871 & 52666 & 24 &  & 
 0.02 & & & 0.02 & & & & & \\
  & (3, 5) & 328.4 & 18390 & 9 &  & 
  0.02 & & & 0.02 & & & & & \\
 & (4, 12) & 4909 & 37083 & 104 &  & 
 0.03 & & & 0.02 & & & & & \\
 
 \hline \hline
 
 \textit{Con.}& (2, 2) & 0.09 & 4 & 4 & 4.34 & 
 0.01 & 1.17 & 104 & 0.01 & 1.17 & 104 & 107 & 129 & 102\\
 
  \hline \hline
  
 \textit{Coo.}& (3, 5) & 2.60 & 161 & 1 & 3.56 & 
 0.01  & 1.21 & 171 & 0.01 & 1.20 & 168 & 168 & 168 & 173\\
 
 \hline \hline
 
\multirow{2}{*}{\textit{Usb}} & (2, 3) & 0.16 & 8 & 8 & \multirow{2}{*}{3.93} & 
0.01 & \multirow{2}{*}{1.80} & \multirow{2}{*}{302} & 0.01 & \multirow{2}{*}{1.76} & \multirow{2}{*}{304} & \multirow{2}{*}{329} & \multirow{2}{*}{313} & \multirow{2}{*}{358} \\ 
&  (3, 5) & 342.2 & 5638 & 32 &  & 
0.01 & & & 0.01 & & & & & \\

 \hline \hline
 
  \multirow{2}{*}{\textit{St.}} &  (8, 8) & 17.9 & 256 & 256 & \multirow{2}{*}{2.86} & 
  0.02 & \multirow{2}{*}{2.86} & \multirow{2}{*}{295} & 0.02 & \multirow{2}{*}{2.86} & \multirow{2}{*}{291} & \multirow{2}{*}{299} & \multirow{2}{*}{298} & \multirow{2}{*}{260} \\ 
 &  (3, 6) & 164.2 & 6138 & 45 & & 
 0.02 & & & 0.02 & & & & & \\
 
 \hline \hline
 
   \multirow{6}{*}{\textit{Syn.}} & (2, 2) & 0.18 & 7 & 2 & 3.82 & 
   0.01 & 1.01 & 106 & 0.01 & 1.03 & 119 & 124 & 129 & 128 \\ 
 & (2, 3) & 1.15 & 53 & 3 & 3.89 & 
 0.01 & 1.95 & 112 & 0.01 & 1.90 & 118 & 118 & 110 & 103 \\
 & (2, 4) & 14.51 & 625 & 3 & 3.72 & 
 0.01 & 1.98 & 113 & 0.01 & 1.98 & 113 & 112 & 115 & 127\\
 & (2, 5) & 59.6 & 2707 & 11 & 3.95 & 
 0.01 & 2.07 & 170 & 0.01 & 2.06 & 167 & 144 & 142 & 152\\
  & (2, 6) & 377.7 & 9042 & 24 & 3.91 & 
  0.02 & 2.14 & 194 & 0.01 & 2.09 & 183 & 184 & 191 & 191 \\  
  & (2, 7) & 3008 & 12290 & 45 & 4.29 & 
  0.02 & 2.40 & 209 & 0.02 & 2.21 & 207 & 247 & 222 & 436\\
  \hline

\end{tabular}

\caption{Empirical evaluation of controller construction (\textit{BA}) and execution performance (\textit{1K,10K...}) of different benchmarks (last group refers to adaptive synthesis). All times are measured in seconds, except for \textit{Cn.} and \textit{Pr.}, which are microseconds.
}
  \label{tabBenchmark}
\end{table*}

}

\begin{abstract}
  \begin{quote}
    Reactive synthesis is the process of using temporal logic
    specifications in \LTL to generate correct controllers, but its use
    has been restricted to Boolean specifications.
    Recently, a Boolean abstraction technique allows to translate
    \LTLt specifications that contain literals in theories into
    equi-realizable \LTL specifications.
    However, no synthesis procedure exists yet.
    In synthesis modulo theories, the system to synthesize receives
    valuations of environment variables in a first-order theory
    $\Theo$ and outputs valuations of system variables from $\Theo$.
    In this paper, we address how to syntheize a full controller using
    a combination of the static Boolean controller obtained from the
    \textit{Booleanized} \LTL specification together with dynamic queries to a solver
    that produces models of a satisfiable existential formulae from $\Theo$.
    %
    This is the first method that realizes reactive synthesis modulo
    theories.
\end{quote}
\end{abstract}


\section{Introduction}

Reactive synthesis is the problem of automatically producing a system
that models a given temporal specification, where the Boolean
variables (i.e., atomic propositions) are split into variables
controlled by the environment and variables controlled by the system.
Realizability is the related decision problem of deciding whether such
a system exists.
These problems have been widely studied~\cite{pnueli89onthesythesis},
specially in the domain of Linear Temporal Logic
(\LTL)~\cite{pnueli77temporal}.
Realizability corresponds to an infinite game where players
alternatively choose the valuations of the Boolean variables they
control.
A specification is realizable if and only if the system has a strategy
such that the specification is satisfied in all plays played according to
the strategy.
The synthesis process is produced from a winning system strategy.
Both reactive synthesis and realizability are decidable for
\LTL~\cite{pnueli89onthesythesis}.
\LTL modulo theories (\LTLt) is the extension of \LTL where Boolean
atomic propositions can be literals from a (multi-sorted) first-order
theory $\Theo$.
Realizability of \LTLt specifications is decidable under certain
conditions over $\Theo$, shown in~\cite{rodriguez23boolean} using a
Boolean abstraction or \textit{Booleanization} method that translates specifications in \LTLt
into equi-realizable \LTL formulae, 
which means that the original specification in \LTLt is realizable 
if and only if the produced Boolean LTL specification is realizable, and vice versa.
Note than an \LTLt reactive specification splits the theory
variables into environment controlled and system controlled variables that can
appear in a single literal, while \LTL Boolean atoms belong fully to
either player.

In this paper, we propose a general method that uses
procedures to dynamically produce outputs as the
results of computing models of existential $\Theo$ formulae.
Concretely, the method we propose statically receives an \LTLt
specification $\varphi$, Booleanizes $\varphi$
using~\cite{rodriguez23boolean} and synthesizes a controller $S$ using
standard methods.
Then, dynamically $S$ is combined with a tool that can produce models
of satisfiable $\Theo$ formulae (e.g., an SMT solver) which
collaborate in tandem at each step of the execution.
To guarantee that the reaction is produced at every step, we require
that $\Theo$ has an efficient procedure to provide models of
existential fragments of $\Theo$.
Our approach does not guarantee termination using semi-decidable
$\Theo$.
We also use an additional component, called \texttt{partitioner}, which
discretizes the environment $\Theo$-input providing a suitable input
for the Boolean controller (but this is computed statically).
To the best of our knowledge, this is the first successful decidable reactive synthesis
procedure for \LTLt specifications.


\section{Preliminaries} \label{sec:prelim}

\subsubsection{Boolean abstraction.}

For this paper, we assume the reader is familiar with LTL \cite{pnueli1977temporalLogic}, $\LTLt$ \cite{rodriguez23boolean} and reactive synthesis \cite{thomas08church}.

The Boolean abstraction procedure takes an input formula
$\varphi_{\mathcal{T}}$ with literals $l_i$ and produces a new
specification
$\varphi_{\mathbb{B}} = \varphi[l_i \leftarrow s_i] \wedge \square
\varphi^{extra}$, where $s_i$ are fresh Boolean variables and
$\varphi^{extra} \in \Boolbb$. 
The core of the algorithm is the additional subformula
$\varphi^{\textit{extra}}$ which uses the freshly introduced variables
$s_i$---controlled by the system---as well as additional Boolean
variables $\overline{e}_k$ controlled by the environment and captures
that, for each possible $\overline{e}_k$, the system has the power to
choose a response among a specific $s_i$.
The extra requirement captures precisely the finite collection of input
decisions of the environment (partitions of the environment space of
valuations) and the resulting (finite) choices of the system to respond
(partitions of the system choices that results in the same Boolean valuations
of the literals).

\subsubsection{Motivating running example.}
As for an example of reactive specifications in LTL$_{\mathcal{T}}$,
let $\square$ be the usual \textit{globally} operator in LTL and
$\Next$ the \textit{next} operator.  Consider
$\phiT = \square (R_0 \wedge R_1)$ as the running example for the paper, where
\[
  R_0 : (x<2) \shortrightarrow \Next(y>1) \hspace{3em}
  R_1 : (x \geq 2) \shortrightarrow (y < x)\]
In $\phiT$, $x \in \Theo$ belongs to the environment and $y \in \Theo$
belongs to the system.
Note that $\phiT$ is not realizable for $\Theo = \ThZ$, since, if at a
given time instant $t$, the environment plays $x=0$, and hence $(x<2)$
holds, then $y$ must be greater than $1$ at time $t+1$. Then, if at
$t+1$ the environment plays $x=2$, then $(x\geq 2)$ holds but there is
no $y$ such that both $(y>1)$ and $(y<2)$.
However, for $\Theo = \ThR$, $\varphi$ is realizable (consider
the system strategy to always play $y=1.5$).
%
%
The Boolean abstraction method transforms $\phiT$ into a purely
Boolean specification $\varphi_{\mathbb{B}}$
that allows to perform automatic \LTL realizability checking. 
For instance, for $\Theo = \ThZ$, the Booleanized version of $\phiT$
is the following:
\[ \varphi_{\mathbb{B}} = \varphi'' \wedge \Always (\phiLegal \Into \phiExtra),
\]
where $\phiLegal$ encodes that $e_0$, $e_1$ and $e_2$ characterize a
partition of the input decisions of the environment
$(e_0 \vee e_1 \vee e_2) \wedge (e_0 \Into \neg (e_1\wedge e_2)) \wedge
(e_1 \Into \neg (e_0\wedge e_2)) \wedge (e_2 \Into \neg(e_0\wedge e_1))$.
Also
$\varphi'' = (s_0 \shortrightarrow \Next s_1) \wedge (\neg s_0
\shortrightarrow s_2)$ is a direct translation of $\phiT$, where $s_0$
abstracts the literal $(x<2)$, $s_1$ abstracts $(y>1)$ and $s_2$ abstracts
$(y<x)$.
Solely replacing literals with fresh system variables over-approximates
the power of the system, therefore we need an additional formula $\phiExtra$
that encodes the original power of each player in $\phiT$:

\newcommand{\Cone}{s_{01\overline{2}}}
\newcommand{\Ctwo}{s_{0\overline{1}2}}
\newcommand{\Cthree}{s_{0\overline{1}\overline{2}}}
\newcommand{\Cfour}{s_{\overline{0}12}}
\newcommand{\Cfive}{s_{\overline{0}1\overline{2}}}
\newcommand{\Csix}{s_{\overline{0}\overline{1}2}}

 \begin{align*}
 \varphi^\textit{extra} :
   \begin{pmatrix}
     \begin{array}{lrcl}
\phantom{\wedge}& \big( e_{0} & \shortrightarrow &  \Cone \vee \Ctwo \vee \Cthree \big)\\[0.27em]
\wedge  & \big(e_{1} &\shortrightarrow & \Cfive \vee \Csix\big) \\[0.27em]
\wedge & \big(e_2 & \shortrightarrow & \Cfour \vee \Cfive \vee \Csix \big)
     \end{array}
   \end{pmatrix},
 \end{align*}
 where $e_0, e_1, e_2 \in\mathbb{B}$ belong to the environment and where $\Cone=(s_{0} \wedge s_{1} \wedge \neg s_{2})$, $\Ctwo=(s_{0} \wedge \neg s_{1} \wedge s_{2})$, $\Cthree=(s_{0} \wedge \neg s_{1} \wedge \neg s_{2})$, $\Cfour=(\neg s_{0} \wedge s_{1} \wedge s_{2})$, $\Cfive=(\neg s_{0} \wedge s_{1} \wedge \neg s_{2})$ and $\Csix=(\neg s_{0} \wedge \neg s_{1} \wedge s_{2})$, where $s_0, s_1,s_2 \in\mathbb{B}$ belong to the system.
 Sub-formulae $\Cone, \Ctwo, \Cthree, \Cfour, \Cfive$ and $\Csix$
 represent the \textit{choices} of the system, that is, given a
 decision $e_k$ of the environment, the system can \textit{react} with
 one of the choices $c_i$ in the disjunction implied by $e_k$.
 Note that $\phiLegal$ encodes that $e_0,e_1,e_2$ is a (finite)
 partition in the domain of the (infinite) valuations of the
 environment, where $e_0$ abstracts its decision $x$ such that
 $(x < 2)$, $e_1$ represents $x$ such that $(x = 2)$ and $e_2$
 represents $(x > 2)$.
%
Note that if the considered $\Theo$ is different, $\varphi_{\mathbb{B}}$ may also differ.
%


\section{Description of the Approach} \label{sec:descrip}

For synthesis modulo theories it is not enough to synthesize a
controller for the Booleanized \LTLt specifications, because the
actual controller will receive inputs in $\Theo$ from the environment and produce
outputs from complex values in $\Theo$. 
For instance, consider a specification where the environment controls
an integer variable $x$ and the system controls an integer variable $y$ in the
specification $\PhiT = \Always (y>x)$.

\begin{figure}[t!]
  \centering
  \includegraphics[scale=0.48]{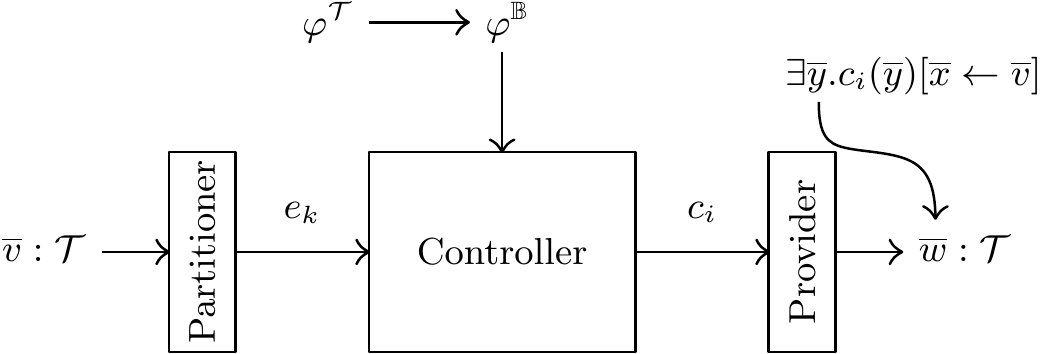}
  \caption{The dynamic synthesis architecture.}
  \label{figMimickStrategy}
\end{figure}

In this paper we propose general alternative approach, shown in
Fig.~\ref{figMimickStrategy}, which we call \textit{dynamic \LTLt
  synthesis}.
This method consists on computing statically a Boolean controller for
$\PhiB$ (which has been Booleanized from $\PhiT$), and dynamically
combine it with a method to provide models from formulae in $\Theo$.
At runtime, at each instant of time, (1) given the valuations
$[\xs \leftarrow \vs]$ of the environment (where $\vs$ are actual input
values for each environment variable $x \in \Theo$), then (2) the
\texttt{partitioner} discretizes this valuation generating a Boolean
input for the Boolean controller; (3) the controller responds with a
choice $c_i \in \mathbb{B}$ (which corresponds to a verdict on the Boolean valuations
of literals in the formula).
Our controller still needs to produce actual values of the output
variables that make the verdict of the literals be as in $c_i$, for which a formula of
the form $\exists \ys \suchThat c^{\Theo}_i(\ys)$ is generated
(where $c^{\Theo}_i(\ys)$ is the $\Theo$ formula that contains one
conjunction per literal, and the input variables replaced by their
values).
This formula represents all the values that the system controls, that
result in the choice $c_i$ that the Boolean controller has output.
By the correctness of the Booleanization process this formula must be
satisfiable.
Stage (4), called \texttt{provider}, uses an SMT solver to produce a
model $\overline{w}$ of $ \exists \ys \suchThat c^{\Theo}_i(\ys)$ so
$[\ys \leftarrow \overline{w}]$ will guarantee the original specification
$\PhiT$. 
Note that we replace $\xs$ by the input valuation $\vs$ in $c^{\Theo}_i(\ys)$,
so $c^{\Theo}_i(\ys)$ only has $\ys$ as variables.

\subsubsection{Execution}

We now illustrate using the running example $\phiT$ how the
dynamic approach behaves in practise.
Specification $\phiT$ is unrealizable for $\ThZ$, but a slight
modification makes it realizable.
If we replace $(y<x)$ with $(y \leq x)$ we obtain
$\phiT' = \square (R_0 \wedge R_1')$, where:
\[ R_0 : (x<2) \shortrightarrow \Next(y>1) \hspace{4em}
  R_1' : (x \geq 2) \shortrightarrow (y \leq x)
\]
Specification $\phiT'$ is realizable in $\ThZ$ (consider the strategy
of the system to always play $y=2$).
The Booleanized version of $\phiT'$ is
$ \phiB' = \varphi'' \wedge \Always([(e_0 \vee e_1) \wedge (e_0 \leftrightarrow
\neg e_1)] \Into \phiExtraP)$, where
$\varphi'' = (s_0 \shortrightarrow \Next s_1) \wedge (\neg s_0
\shortrightarrow s_2)$ and $\phiExtraP$ is:
\begin{align*}
  \phiExtraP:
  \begin{pmatrix}
    \begin{array}{lrcl}
      \phantom{\wedge}& \big(e_0 & \Into & \big( \Cone \vee \Ctwo \big) \\[0.27em]
      \wedge & \big(e_1 &\Into &\big(\Cfour \vee \Cfive \vee \Csix\big)
    \end{array}
  \end{pmatrix},
\end{align*}
where $e_0, e_1 \in\mathbb{B}$ belong to the environment and represent
$(x<2)$ and $(x \geq 2)$, respectively. 
Note that in $\phiB'$ there are no separated $e_k$ for $(x=2)$ and $(x>2)$.
We show a concrete execution in Tab.~\ref{tabExample}, where we see
how the $\Theo$-controller responds to a few $\Theo$-inputs.
For instance, in the first step, the input $x = 4$ is discretized into
the Boolean decision $e_1$ which is passed to the Boolean controller.
The controller responds $\Cfour = \neg s_0 \wedge s_1 \wedge s_2$
to this input, which is translated into
$\Cfour^{\Theo} = \neg (x<2) \wedge (y>1) \wedge (y \leq x)$.
Then, \texttt{provider} substitutes valuation $[x \leftarrow 5]$ in
$\Cfour^{\Theo}$, and solves
$\exists y. \Cfour^{\Theo}(y) [x \leftarrow 5]$, i.e.,
$\exists y. \neg (5<2) \wedge (y>1) \wedge (y \leq 5)$ which is
guaranteed to succeed. 
A possible model is $y=2$.

\newcommand{\Col}[1]{\colorbox{Gray}{$#1$}}

\begin{table}[b]
\centering
  \begin{tabular}{|@{$\;$}c@{$\;$}|@{}c@{}|@{}c:c@{}|c|} \hline
    {\small Step} & $x$ & $\overline{e}$ & $\overline{c}$ & $\ys$ \\
    \hline
    1 & $..., 3, \Col{4}, 5, ...$ & $e_0, \Col{e_1}$ & $\Col{\Cfour}, \Cfive, \Csix$ & $3, \Col{2}, ...$ \\
    
    2 & $..., 3, \Col{4}, 5, ...$ & $e_0, \Col{e_1}$ & $\Col{\Cfour}, \Cfive, \Csix$ & $3, \Col{2},...$ \\
    
    3 & $.., 0, \Col{1}, 2, ...$ & $\Col{e_0},e_1$ & $\Cone, \Col{\Ctwo}$ & $\Col{1}, 0,...$ \\
    
    4 & $..,-1, \Col{0}, 1, ...$ & $\Col{e_0},e_1$ & $\Col{\Cone}, \Ctwo$ & $\Col{2}, 3,..$ \\
    
    5 & $..\Col{2}, 3, 4,\ldots$ & $e_0, \Col{e_1}$ & $\Col{\Cfour}, \Cfive, \Csix$ & $\Col{2}$ \\
  \end{tabular}
  \caption{Modified running example $\phiT'$ executed for 5 steps. Gray colour indicates the selected values among the infinitely many $\xs$ and $\ys$ or the finitely many $\overline{e}_k$ and $c_i$ options.}
  \label{tabExample}
\end{table}

\newcommand{\ek}{\ensuremath{\overline{e}_k}\xspace}


\section{Related Work and Conclusions} \label{sec:conclusion}

\subsubsection{Related Work.}

Recently, \cite{rodriguez23boolean} introduced $\LTLt$, and showed that the
realizability problem for $\LTLt$ is decidable via a Boolean
abstraction technique.
We extended this approach here to full reactive synthesis modulo
theory.
Alternative definitions for LTL modulo theories~\cite{geatti22linear}
have been developed for finite traces, but allowing temporal operators
within predicates, which makes the logic undecidable.  Similar
undecidability is reported
in~\cite{faranKupferman2022LTLwithArithmeticApplicationsR}.
Other approaches (e.g., \cite{demriDSouza2002automataTheoreticApproachConstraintLTL,cheng2013numerical,azadehKincaid2017strategySynthesisLinearArithmetic}) 
restrict expressivity whether temporal-wise, theory-wise or both.

Some works (e.g.,
\cite{katisETAL2016synthesisAssumeGuaranteeContracts,katisETAL2018validityGuidedSynthesis,gacekETAL2015towardsRealizabilityCheckingContracts,walkerRyzhyk2014predicateAbstractionReactiveS}
consider synthesis or realizability of first-order theories, but none
of them offers termination guarantees and they only consider some
temporal fragments.
Our approach guarantees termination of the computation of the controller 
if the used theory is decidable in the $\exists^*\forall^*$ fragment 
and runtime guarnatees termination in each timestep if the SMT solver supports the theory.
Moreover, all these approaches above adapt one specific technique and
implement it in a monolithic way, whereas Boolean abstraction allows
us the construct the general dynamic architecture, since it
generates an equi-realizable (Boolean) LTL specification.
Note that our dynamic approach benefits from all advantages of
using synthesis from Boolean abstractions and is fully automatic
(unlike \cite{walkerRyzhyk2014predicateAbstractionReactiveS}).

Temporal Stream Logic
(TSL)~\cite{finkbeinerETAL2017temporalStreamLogicSynthesisBeyondBools}
extends LTL with complex data that can be related accross time and
\cite{finkbeinerETAL2021temporalStreamLogicModuloTheories,maderbacherBloem2021reactiveSynthesisModuloTheoriesAbstraction,wonhyukETAL2022canSynthesisSyntaxBeFriends}
use extensions of TSL to theories.
Again, realizability (and thus synthesis) is undecidable in all
these works.
In comparison, our Boolean abstraction cannot relate values accross
time but provides a decidable synthesis procedure.
Also, TSL is undecidable already for safety, the theory of equality
and Presburger arithmetic.
More precisely, TSL is only known to be decidable for three fragments
(see Thm. 7 in
\cite{finkbeinerETAL2021temporalStreamLogicModuloTheories}).

\subsubsection{Conclusion.}

We have studied the problem of $\LTLt$ synthesis which is more
challenging than $\LTLt$ realizability modulo theories, since synthesis
implies computing a system that receives valuations in $\Theo$ and
provides valuations in $\Theo$.
We propose an \textit{dynamic} approach that first discretizes the
input from the environment, then uses a Boolean controller synthesized
from the Booleanized specification of $\LTLt$, and finally produces a
reaction using a procedure that provides models of existential
formulae of $\Theo$.


\bibliographystyle{aaai}
\bibliography{references}

\begin{thebibliography}{}

\bibitem[\protect\citeauthoryear{Cheng and Lee}{2013}]{cheng2013numerical}
Cheng, C., and Lee, E.~A.
\newblock 2013.
\newblock Numerical {LTL} synthesis for cyber-physical systems.
\newblock {\em CoRR} abs/1307.3722.

\bibitem[\protect\citeauthoryear{Choi \bgroup et al\mbox.\egroup
  }{2022}]{wonhyukETAL2022canSynthesisSyntaxBeFriends}
Choi, W.; Finkbeiner, B.; Piskac, R.; and Santolucito, M.
\newblock 2022.
\newblock Can reactive synthesis and syntax-guided synthesis be friends?
\newblock In {\em Proc. of the 43rd {ACM} {SIGPLAN} Int'l Conf. on Programming
  Language Design and Implementation (PLD'22)},  229--243.
\newblock {ACM}.

\bibitem[\protect\citeauthoryear{Demri and
  D'Souza}{2007}]{demriDSouza2002automataTheoreticApproachConstraintLTL}
Demri, S., and D'Souza, D.
\newblock 2007.
\newblock An automata-theoretic approach to constraint {LTL}.
\newblock {\em Inf. Comput.} 205(3):380--415.

\bibitem[\protect\citeauthoryear{Faran and
  Kupferman}{2018}]{faranKupferman2022LTLwithArithmeticApplicationsR}
Faran, R., and Kupferman, O.
\newblock 2018.
\newblock {LTL} with arithmetic and its applications in reasoning about
  hierarchical systems.
\newblock In {\em Proc. of the 22nd International Conference on Logic for
  Programming, Artificial Intelligence and Reasoning, ({LPAR-22.} ), Awassa,
  Ethiopia, 16-21 November 2018}, volume~57 of {\em EPiC Series in Computing},
  343--362.
\newblock EasyChair.

\bibitem[\protect\citeauthoryear{Farzan and
  Kincaid}{2018}]{azadehKincaid2017strategySynthesisLinearArithmetic}
Farzan, A., and Kincaid, Z.
\newblock 2018.
\newblock Strategy synthesis for linear arithmetic games.
\newblock {\em Proc. {ACM} Program. Lang.} 2({POPL}):61:1--61:30.

\bibitem[\protect\citeauthoryear{Finkbeiner \bgroup et al\mbox.\egroup
  }{2019}]{finkbeinerETAL2017temporalStreamLogicSynthesisBeyondBools}
Finkbeiner, B.; Klein, F.; Piskac, R.; and Santolucito, M.
\newblock 2019.
\newblock Temporal stream logic: Synthesis beyond the {B}ools.
\newblock In Dillig, I., and Tasiran, S., eds., {\em Proc. of the 31st Int'l
  Conf. on Computer Aided Verification (CAV'19), Part {I}}, volume 11561 of
  {\em LNCS},  609--629.
\newblock Springer.

\bibitem[\protect\citeauthoryear{Finkbeiner, Heim, and
  Passing}{2022}]{finkbeinerETAL2021temporalStreamLogicModuloTheories}
Finkbeiner, B.; Heim, P.; and Passing, N.
\newblock 2022.
\newblock Temporal stream logic modulo theories.
\newblock In {\em Proc. of the 25th Int'l Conf. on Foundations of Software
  Science and Computation Structures (FOSSACS'22)}, volume 13242 of {\em LNCS},
   325--346.
\newblock Springer.

\bibitem[\protect\citeauthoryear{Gacek \bgroup et al\mbox.\egroup
  }{2015}]{gacekETAL2015towardsRealizabilityCheckingContracts}
Gacek, A.; Katis, A.; Whalen, M.~W.; Backes, J.; and Cofer, D.~D.
\newblock 2015.
\newblock Towards realizability checking of contracts using theories.
\newblock In {\em Proc. of the 7th International Symposium {NASA} Formal
  Methods (NFM'15)}, volume 9058 of {\em LNCS},  173--187.
\newblock Springer.

\bibitem[\protect\citeauthoryear{Geatti, Gianola, and
  Gigante}{2022}]{geatti22linear}
Geatti, L.; Gianola, A.; and Gigante, N.
\newblock 2022.
\newblock Linear temporal logic modulo theories over finite traces.
\newblock In {\em Proc. of the 31st International Joint Conference on
  Artificial Intelligence, ({IJCAI} 2022), Vienna, Austria, 23-29 July 2022},
  2641--2647.
\newblock ijcai.org.

\bibitem[\protect\citeauthoryear{Katis \bgroup et al\mbox.\egroup
  }{2016}]{katisETAL2016synthesisAssumeGuaranteeContracts}
Katis, A.; Fedyukovich, G.; Gacek, A.; Backes, J.~D.; Gurfinkel, A.; and
  Whalen, M.~W.
\newblock 2016.
\newblock Synthesis from assume-guarantee contracts using skolemized proofs of
  realizability.
\newblock {\em CoRR} abs/1610.05867.

\bibitem[\protect\citeauthoryear{Katis \bgroup et al\mbox.\egroup
  }{2018}]{katisETAL2018validityGuidedSynthesis}
Katis, A.; Fedyukovich, G.; Guo, H.; Gacek, A.; Backes, J.; Gurfinkel, A.; and
  Whalen, M.~W.
\newblock 2018.
\newblock Validity-guided synthesis of reactive systems from assume-guarantee
  contracts.
\newblock In {\em Proc. of the 24th Int'l Conf. on Tools and Algorithms for the
  Construction and Analysis of Systems, (TACAS'18), Part {II}}, volume 10806 of
  {\em LNCS},  176--193.
\newblock Springer.

\bibitem[\protect\citeauthoryear{Maderbacher and
  Bloem}{2022}]{maderbacherBloem2021reactiveSynthesisModuloTheoriesAbstraction}
Maderbacher, B., and Bloem, R.
\newblock 2022.
\newblock Reactive synthesis modulo theories using abstraction refinement.
\newblock In {\em 22nd Formal Methods in Computer-Aided Design, (FMCAD'22)},
  315--324.
\newblock {IEEE}.

\bibitem[\protect\citeauthoryear{Pnueli and
  Rosner}{1989}]{pnueli89onthesythesis}
Pnueli, A., and Rosner, R.
\newblock 1989.
\newblock On the synthesis of an asynchronous reactive module.
\newblock In {\em Proc. of the 16th Int'l Colloqium on Automata, Languages and
  Programming (ICALP'89)}, volume 372 of {\em LNCS},  652--671.
\newblock Springer.

\bibitem[\protect\citeauthoryear{Pnueli}{1977a}]{pnueli77temporal}
Pnueli, A.
\newblock 1977a.
\newblock The temporal logic of programs.
\newblock In {\em Proc. of the 18th IEEE Symp. on Foundations of Computer
  Science (FOCS'77)},  46--67.
\newblock IEEE CS Press.

\bibitem[\protect\citeauthoryear{{Pnueli}}{1977b}]{pnueli1977temporalLogic}
{Pnueli}, A.
\newblock 1977b.
\newblock The temporal logic of programs.
\newblock {\em Proceedings of the 18th Annual Symposium on Foundations of
  Computer Science (FOCS)}  46–57.

\bibitem[\protect\citeauthoryear{Rodriguez and
  S\'{a}nchez}{2023}]{rodriguez23boolean}
Rodriguez, A., and S\'{a}nchez, C.
\newblock 2023.
\newblock Boolean abstractions for realizabilty modulo theories.
\newblock In {\em Proc. of the 35th International Conference on Computer Aided
  Verification (CAV'23)}, volume 13966 of {\em LNCS}.
\newblock Springer, Cham.

\bibitem[\protect\citeauthoryear{Thomas}{2008}]{thomas08church}
Thomas, W.
\newblock 2008.
\newblock Church's problem and a tour through automata theory.
\newblock In {\em In Pillars of Computer Science, Essays Dedicated to Boris
  (Boaz) Trakhtenbrot on the Occasion of His 85th Birthday}, volume 4800 of
  {\em LNCS},  635--655.
\newblock Springer.

\bibitem[\protect\citeauthoryear{Walker and
  Ryzhyk}{2014}]{walkerRyzhyk2014predicateAbstractionReactiveS}
Walker, A., and Ryzhyk, L.
\newblock 2014.
\newblock Predicate abstraction for reactive synthesis.
\newblock In {\em Proc. of the 14th Formal Methods in Computer-Aided Design,
  ({FMCAD} 2014), Lausanne, Switzerland, October 21-24, 2014},  219--226.
\newblock {IEEE}.

\end{thebibliography}

\vfill


\end{document}